\documentclass[manuscript,nolinenumbers]{aastex701}

\usepackage{bm}
\usepackage{amsmath}
\usepackage{color}
\usepackage{float}

\shorttitle{Quasi-biennial and Rieger-type periods}
\shortauthors{Stefani et al.}
\graphicspath{{./}{figures/}}

\begin{document}

\title{From the QBO to Rieger-type periods: Explaining extreme solar events by tidally triggered magneto-Rossby waves}

\author[0000-0002-8770-4080]{Frank Stefani}
\affiliation{Institute of Fluid Dynamics, Helmholtz-Zentrum Dresden-Rossendorf, Bautzner Landstrasse 400, 01328 Dresden, Germany}
\email{F.Stefani@hzdr.de}

\author[0000-0001-9892-9309]{Gerrit M. Horstmann}
\affiliation{Institute of Fluid Dynamics, Helmholtz-Zentrum Dresden-Rossendorf, 
Bautzner Landstrasse 400, 01328 Dresden, Germany}
\email{G.Horstmann@hzdr.de}

\author[0000-0002-6189-850X]{George Mamatsashvili}
\affiliation{Institute of Fluid Dynamics, Helmholtz-Zentrum Dresden-Rossendorf, 
Bautzner Landstrasse 400, 01328 Dresden, Germany}
\affiliation{Abustumani Astrophysical Observatory, Mount Kanobili, Abustumani 0301, Georgia}
\email{G.Mamatsashvili@hzdr.de}

\author[0009-0007-0076-6531]{Tom Weier}
\affiliation{Institute of Fluid Dynamics, Helmholtz-Zentrum Dresden-Rossendorf, Bautzner Landstrasse 400, 01328 Dresden, Germany}
\email{T.Weier@hzdr.de}



\begin{abstract}
It has recently been demonstrated that the occurrence of extreme solar events is governed by a phase-stable period close to 1.723 years (629.29\,days). According to our previously developed tidally-synchronized solar dynamo model, this is one of the periods resulting from two-planet spring tides exerted by the tidally dominant planets Venus, Earth and Jupiter, which trigger magneto-Rossby waves in the solar tachocline. Overall, the squared sum of these three waves incorporates nine periods ranging from 59 to 629 days. 
Using a standard Rayleigh test, we demonstrate that all nine of these periods can be identified in the data on extreme solar events. For the latter, we use a combination of 72 ground-level enhancement (GLE) events with various X-flare data, the number of which increases when the threshold  
is successively lowered from X10 to X1. To identify the spectral peaks, 
it is necessary to analyze the sidebands that emerge due to the modulation with the 
longer-term periods such as the 11-year Schwabe cycle and the third harmonics 
of the Hale cycle. In this framework, the seminal Rieger period of 154 days is
re-interpreted as the lower sideband of 146\,days which is the quarter of 
the synodic period of Venus and Earth.
\end{abstract}


\section{Introduction} \label{sec:intro}

A key finding of \cite{Velasco2018} was that the 57 ground-level enhancement (GLE) events between 1955 and 2017 occurred predominantly during the positive phase of an oscillation with a period of 1.73 years. The emergence of such a specific value is highly surprising, given that 
the quasi-biennial oscillation (QBO) of solar activity is generally 
considered to lie somewhere between 0.6 and 4 years \citep{Bazilevskaya2014}.
In fact, the observed phase stability over around five solar cycles is 
almost impossible to explain without assuming the existence of an external 
``clock'' of some kind. But what could this clock be?

While attributing the QBO to magneto-Rossby waves has a relatively 
long tradition that traces back to \cite{Zaqarashvili2010}, the 
specific number of 1.73\,years found by \cite{Velasco2018} prompted us to tentatively 
associate the ``clock'' with the periodic tidal action of the orbiting planets.
Based on the finding by \cite{Horstmann2023} that the tidal forces exerted by the 
planets can trigger magneto-Rossby waves at the solar tachocline with velocities 
in the order of m/s or larger, \cite{Stefani2025} had shown that 
the two-planet spring tides of Venus-Jupiter (determined by half of 
their synodic period, i.e. $0.5 P_{\rm VJ}=118.50$\,days), 
Earth-Jupiter ($0.5 P_{\rm EJ}=199.44$\,days), and Venus-Earth 
($0.5 P_{\rm VE}=291.97$\,days) lead to a dominant beat 
period of 629.29 days, or 1.723 years, which is in remarkable 
agreement with the observation 
of \cite{Velasco2018}. Later
\cite{Stefani2026} showed that 
not only the GLE events, but also X10-flares are
dominated by this particular period.

 The attribution of the QBO to three tidally triggered magneto-Rossby waves is a relatively recent development that followed the earlier identification of the 11.07-year Schwabe cycle as a secondary beat period of the same waves.
 After a decade of increasingly complex synchronization 
studies \citep{Stefani2016,Stefani2019,Stefani2021,Klevs2023}, the 
current theory now provides a plausible and self-consistent explanation
of the observed 
periods of the QBO and the Schwabe/Hale cycle \citep{Stefani2025}.
It also accurately describes the periods of the 
Gleissberg and Suess-de Vries cycles
(as observed in the Lake Lisan sediments by \cite{Prasad2004})
if some  sort of spin-orbit coupling effect due to the barycentric 
motion of the Sun, related to the
19.86-year synodic period of Jupiter and Saturn, is 
additionally implemented in the underlying 
solar dynamo model \citep{Stefani2024}.

In view of this situation, it seems worthwhile to also investigate the remaining eight periods resulting from squaring the sum of the three waves, apart from the QBO.
The rationale behind considering this square is based on the assumption 
that the most probable action of the magneto-Rossby waves on the solar dynamo is related to a certain quadratic form of the waves, might this be a wave-induced $\alpha$-effect in the tachocline, or some Reynolds 
stress therein causing additional zonal or meridional circulation.

Of the nine periods arising from squaring the sum 
of the three waves the 629.29-day period 
is just
the longest one. As shown by  \cite{Stefani2025}, the square 
of the three waves
comprises the quarters of the synodic periods, i.e. 
\begin{eqnarray}
P_{59} = 0.25 P_{\rm VJ} = 59.25\,{\rm days}, \; P_{100} = 0.25 P_{\rm EJ} = 99.72\,{\rm days}, \; P_{146} = 0.25 P_{\rm VE} = 145.99\,{\rm days}  \;  \nonumber
\end{eqnarray}
(note that in this work we replace the formal indices $1...9$ of 
the periods $P$, as used
in \cite{Stefani2025}, by the more meaningful (approximate) periods in days). 
Six more periods emerge from the composition of two periods:
\begin{eqnarray}
   &P_{118} = \frac{0.5P_{\rm VE}P_{\rm EJ}}{P_{\rm VE}+P_{\rm EJ}} = 118.5\,{\rm days} \qquad P_{629} = \frac{0.5P_{\rm VE}P_{\rm EJ}}{P_{\rm VE}-P_{\rm EJ}} = 629.29\,{\rm days} \nonumber \\	
  &P_{84} = \frac{0.5P_{\rm VE}P_{\rm VJ}}{P_{\rm VE}+P_{\rm VJ}} = 84.29\,{\rm days} \qquad  P_{199} = \frac{0.5P_{\rm VE}P_{\rm VJ}}{P_{\rm VE}-P_{\rm VJ}} = 199.43\,{\rm days} \nonumber \\
    &P_{74} = \frac{0.5P_{\rm EJ}P_{\rm VJ}}{P_{\rm EJ}+P_{\rm VJ}} = 74.33\,{\rm days} \qquad P_{292} = \frac{0.5P_{\rm EJ}P_{\rm VJ}}{P_{\rm EJ}-P_{\rm VJ}} = 291.96\,{\rm days} \; . \nonumber 
\end{eqnarray}
Interestingly, these periods are not all independent from the initial synodic periods. Specifically, we can identify
\begin{eqnarray}
 P_{59}=0.5 P_{118} = 0.25 P_{\rm VJ}  \qquad P_{100}=0.5 P_{199} =  0.25 P_{\rm EJ} \qquad P_{146}=0.5 P_{292} = 0.25 P_{\rm VE} \;. \nonumber
 \label{eq:periods}
\end{eqnarray}
In summary, the joint action of the three Rossby waves characterized by the quadratic form contains the three spring-tide periods, their half-periods and three composed periods, among them the important QBO-period $P_{629}$.

Two remarks are in order here: first, as shown in
\cite{Stefani2025}, only the periods $P_{199}$, $P_{292}$, and $P_{629}$ remain when computing the {\it axisymmetric} part 
of the squared sum of the three waves.
While this restriction may indeed have consequences for identifying dynamo-relevant periods (including the secondary 11.07-year beat), it is likely that non-axisymmetric contributions also play a role in launching flux tubes, as suggested by the concept of {\it tachocline nonlinear oscillations} \citep{Dikpati2018}.
Second, among all periods $P_{146}=145.99$\,days is the closest
one to the important {\it Rieger period}, which was first 
identified by \cite{Rieger1984} as 154 days, but whose precise value is meanwhile known to depend on the cycle and the considered solar 
hemisphere (see Figures 2-4 in \cite{Gurgenashvili2021}).

 This paper will therefore examine whether the remaining eight periods discussed above can be identified in various data sets of extreme solar events, in addition to the 629-day period already established.  To that end, we will once again merge the 72 available GLE events with various X-class flare data.  
For selecting the latter, we will use various thresholds 
ranging from extreme X10 flares to less energetic X1 flares. 
In terms of statistical tools, we will rely solely on the standard Rayleigh test to identify the dominant phase-stable periods.

We will demonstrate that most of the eight additional periods can be clearly observed in the data, although one must always carefully consider the side bands that emerge due to the simultaneous presence of the longer-term periods, including the dominant Schwabe cycle, the Hale cycle, and its third harmonic. The Rieger-type period $P_{146}$ will be discussed in particular detail.

\section{Data}
In order to test for any underlying periodicity in solar activity events, we will analyze various combinations of datasets with a decreasing threshold of event intensity.
Motivated by \cite{Velasco2018}, and following
\cite{Stefani2025,Stefani2026}, 
we will again use the 72 GLE events that were observed 
between February 1956 and November 2025.
These data, obtained from https://gle.oulu.fi and documented in Table 1 of 
\cite{Stefani2026}, are shown as the green open 
circles in Figure 1a. 

The second dataset comprises X-class flares of different intensities.
The idea behind using different thresholds is that it is a priori 
not clear which choice will best identify the most relevant periods.
On the one hand, accepting a lower X-flare threshold increases the amount 
of X-flare data available, which could help identify 
peaks in the Rayleigh test.
On the other hand, a lower threshold might include  a larger number of weaker solar events, which
are not necessarily connected with the action of magneto-Rossby waves at the tachocline.

We begin with X10+ flares (or S-class flares), which were obtained from \cite{Tan2025} and previously documented in Table 2 of \cite{Stefani2026}. They are depicted as black open circles in Figure 1a.
The red open circles show the merger of these 37 events with the 72 GLE events. 
From this point onward, we will be using increasingly lower-intensity X-flare data from the GOES Solar Flare Catalog (1975-2026). These data were basically obtained 
from https://rudimirz.com/flares, while the intensity of some of the higher events 
were ``upgraded'' to X10+ if this was suggested by the Table 1 of \cite{Tan2025}.  
The corresponding open circles 
in Figure 1(a) comprise 49 of X6+ flares (violet), 137 of X3+ flares (light blue),
252 of X2+ flares (orange), and 569 of X1+ flares (dark blue).

Figure 1a shows also the 13-month averaged sunspot numbers
(yellow curve) as obtained from www.sidc.be/SILSO/infosnmstot .
As in \cite{Stefani2026a}, this smoothed dataset is considered 
a good proxy
for the magnetic field prevailing at the tachocline during (or shortly before) 
the extreme solar events to be considered.
Not surprisingly, both GLE events and flares occur preferentially
(although not exclusively) around the maxima of the solar cycle.

As is well known, GLE are often connected with
strong flares, which explains the partial overlap of
the green and the black open circles in Figure 1(a). 
The inclusion  of the GLE events has, however, the advantage of 
providing a longer data set which helps to improve the accuracy of 
the long-term periods.

\begin{figure}[t]
  \centering
\includegraphics[width=0.99\textwidth]{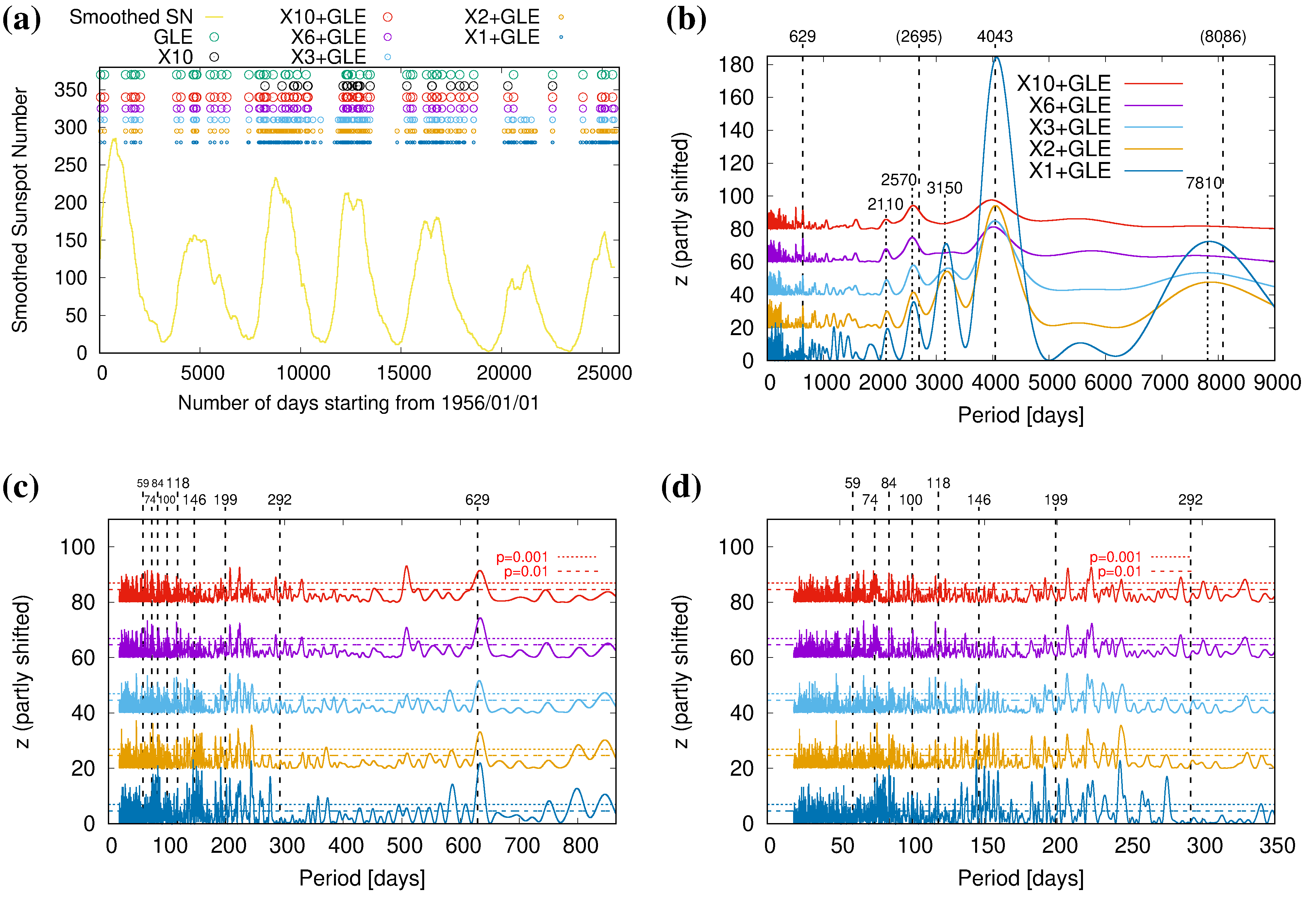}
  \caption{
 Overview of the utilized data and the spectra 
 emerging from the Rayleigh test. (a) Smoothed sunspot number (solid yellow line) for the time interval between 1956 January 1 and 2026 February 1, together with various 
 collections of extreme solar events shown as open circles: 72 GLE events only (green), 
 37 X10+ flares (black), merger
 of 72 GLE events with 37 X10+ flares (red), with 49 X6+ flares  (violet), with 137 X3+  flares (light blue), with 252 X2+ flares (gold), and with 569 X1+ flares (dark blue).
 (b) $z$-values of the Rayleigh test as a function of the period (between 20 and 9000 days) for the
 five combined data sets of (a). For better visibility, the $z$-values are partly upward-shifted by multiples of 20. The numbers on top of each panel denote the theoretical predictions of 629 days (1.723 years) for the QBO, 4043 days (11.07 years) for the Schwabe period $P_S$, and 8086 days (22.14 years) for the Hale period
 $P_{H}$. The value in parentheses, $P_T=2695$\,days, 
 would correspond to the third harmonics of the Hale cycle.
 While the observed maxima 
 of the $z$-value curve lie indeed very close to the QBO and the Schwabe cycle, there is a significant shift of the Hale cycle to the observed period 
 $P^*_{H}=7750$\,days, and of its third harmonics to $P^*_T=2570$\,days.
 Two further peaks, at $P^*_Q=2100$\,days and at $P^*_{I}=3120$\,days, 
 are presently not 
 explained by theory. (c) Zoomed-in version of (b) for periods
 below 850 days. Again, the numbers on top of each panel indicate the theoretical values. The dashed and dotted horizontal lines indicate those 
 $z-$values that correspond to $p$-values of 0.01 and 0.001, respectively. (d) Same as in (c) but zoomed in for periods below 350 days.}
\end{figure}

\section{The Rayleigh test and its main results}

In this section, we will use the standard Rayleigh test to identify the 
dominant phase-stable periods 
associated with the combined GLE and X-flare data sets shown in Figure 1(a). 
According to \cite{Droege1990}, the test statistic $z$ used in the 
Rayleigh test is
defined as
\begin{eqnarray}
z(T)&=&\frac{1}{N} \left[ \left( \sum^N_{i=1} \cos(2 \pi t_i/T) \right)^2+\left( \sum^N_{i=1} \sin(2 \pi t_i/T) \right)^2 \right] \, .
\end{eqnarray}
Here, $t_i$ represents the $i^{\rm th}$ instant of the $N$ extreme solar 
events under consideration, and $T$ is the variable period for which 
values between 20 and 9,000 days will be considered. 
We will focus mainly on those peaks with sufficiently
high values of $z$, which correspond to a low values of $p\approx \exp(-z)$.
In view of the fact that the total time span $T_t$ considered is approximately 
25,000 days (of which the X-flares cover only around 17,000 days), we
should always bear in mind the possible occurrence of 
relative errors in the derived 
period $T$ of the order of $T/T_t$.

\subsection{Overview}
Panels (b-d) of Figure 1 give an overview of the
$z$-values that result from analyses of different data sets.
The colors red, violet, light blue, gold and dark blue
correspond to those of the respective data sets shown in panel (a).
In order to improve visibility, 
the five $z-$value curves are stacked one over the other, 
with relative shifts in the $z$-value of 20. The dashed
and dotted lines indicate the $z$-thresholds of 4.61 and 6.91
which correspond, respectively, to $p$-values of 0.01 and 0.001.
We use here the approximate formula for the single-trial probability 
$p\approx \exp(-z)$ which is
sufficiently accurate for large values of $N$
(for the moment we ignore any statistical intricacies that could be
related with the intrinsic mathematical link between
the nine periods we are focusing on).
All numbers on top of each panel relate to theoretical values 
derived from the Rossby wave theory, while the numbers within the panels 
refer to the observed periods.

The most clearly expressed feature seen in panel (b) is the peak 
very close to the theoretical value of the Schwabe cycle,
$P_S=4043$\,days (i.e. 11.07 years), which becomes increasingly 
pronounced when the threshold for events is lowered, i.e. when 
moving from X10+ to X1+ flares. For low thresholds, the 
concentration of events around the cycle maxima becomes so dense
that $z$-values at $P_S$ reach extremely high
values (up to 180 for the X1 flares).

Interestingly, the Hale cycle also shows up at those low thresholds, 
although at a slightly lower period of $P^*_H=7810$\,days than 
the theoretical value 
of $P_H=8086$\,days (from here on, periods $P$ without
asterisk will denote theoretical values, in contrast to 
periods $P^*$ with asterisk denoting observed values). 
Obviously, for such long periods, we cannot expect high precision.

Above the upper abscissa we have also indicated, although in parentheses, the period
of $P_T=2695$\,days (7.38 years), which would correspond to the  
third harmonics of the {\it theoretical} Hale cycle. Close to one third of the
{\it observed} Hale period, at 
$P^*_T=2570$\,days (7.04 years), we indeed find a well-expressed peak, in 
agreement with the findings
of \cite{Velasco2026}. While the emergence of the 
third harmonics of the
fundamental eigenfrequency is a genuine feature of non-linear
oscillatory dynamos (observed, e.g., in the Riga dynamo experiment 
by
\cite{Gailitis2018}), we should not expect such perfect 
agreement here as in the case of Rossby wave-triggered magnetic 
activity. 

The intermediate peak around $P^*_I=3150$\,days, which only occurs for 
the lowest X-flare thresholds, is close to the average
frequency of the two periods mentioned above. Although we currently 
have no clear explanation for this peak, or the one at $P^*_Q=2110$ days, 
we should not overlook them when investigating side bands of the 
lower-period peaks at a later stage.

Of all those lower-period peaks only the 
QBO-peak at $P_{629}=629.29$\,days (1.723 years) 
is explicitly indicated in panel (b). More details are then visible 
in
panel (c), where all the nine theoretical periods are 
indicated above the upper abscissa. Clearly, the peak 
close to 629\,days stands out as a very pronounced and robust one; it 
seems to become even sharper when lowering the X-flare threshold. 
Generally, however, at (or around) all nine theoretical
periods we observe significant peaks.
A further zoom into the periods below 350 days is provided in 
panel (d). 

\subsection{From the QBO down to the Rieger period}

We begin our detailed analysis with the QBO-period of $P_{629}=629.29$\,days, 
the Venus-Earth spring tide of 
$P_{292}=291.96$\,days, the Earth-Jupiter spring tide of 
$P_{199}=199.43$\,days, and the Rieger-type period of
$P_{146}=145.99$ 
days. To ease the notation, in the following we will
often skip the time unit ``days''.

\begin{table}[h]
\scriptsize
\caption{Analysis of the QBO-period at $P_{629}=629.29$\,days 
and its upper and lower side bands. The upper part of the 
Table shows, in the left column, the observed 
peaks in the upper (u) and lower (l) 
side bands, as indicated in Figure 2(a). These observed periods
are marked by an asterisk. The right column shows the corresponding
theoretical periods (without asterisk) 
of the side peaks as they are expected from the modulation
of the period $P_{629}$ 
with a given longer-term period (here: $P^*_T$, $P^*_I$, or $P_S$).
The lower part of the Table compares the theoretical (without asterisk) or observed (with 
asterisk) periods of $P_{629}$ and $P^*_{T}$ (on the left column) with the  
corresponding inferred values, denoted by a double asterisk on the right column, 
that result from the pair of observed upper and lower side peaks. }
\begin{tabular}{rrrr}
\hline
\hline
Observed side peaks&&Inferred from theory\\
\hline
$P^*_{u,T}=508$&&$P_{u,T}=(P_{629}^{-1}+P_T^{*-1})^{-1}=505.5$ \\
$P^*_{l,T}=848$&&$P_{l,T}=(P_{629}^{-1}-P_T^{*-1})^{-1}=833.3$ \\
$P^*_{l,I}=800$&&$P_{l,I}=(P_{629}^{-1}-P_I^{*-1})^{-1}=786.4$ \\
$P^*_{l,S}=748$&&$P_{l,S}=(P_{629}^{-1}-P_S^{-1})^{-1}=745.3$ \\
\hline
Theory/Observed && Inferred from observed side peaks\\
\hline
$P_{629}=629.3$/$P^*_{629}=633$&&$P_{629}^{**}=2 (P_{u,T}^{*-1}+P_{l,T}^{*-1})^{-1}=635.4$\\
$P_T^*=2570$&&$P_T^{**}=2 (P_{u,T}^{*-1}-P_{l,T}^{*-1})^{-1}=2534$\\
\hline
\end{tabular}
\end{table}

Figure 2(a) shows the $z-$values in the vicinity
of $P_{629}$. As already discussed in  \cite{Stefani2026}, the
observed maximum lies approximately at $P_{629}^*=633$ which is 
0.6\% away from the theoretical value $P_{629}=629.29$.
Such a deviation is well within the characteristic 
relative uncertainty  $\approx 629/25000=2.5$\% that
should be expected in view of the finite interval length of the data.
In any case, for all chosen X-flare 
thresholds the $z$-values are quite large
(leading to $p$-values much below 0.001) which makes 
this central peak highly significant.

\begin{figure}[t]
  \centering
\includegraphics[width=0.99\textwidth]{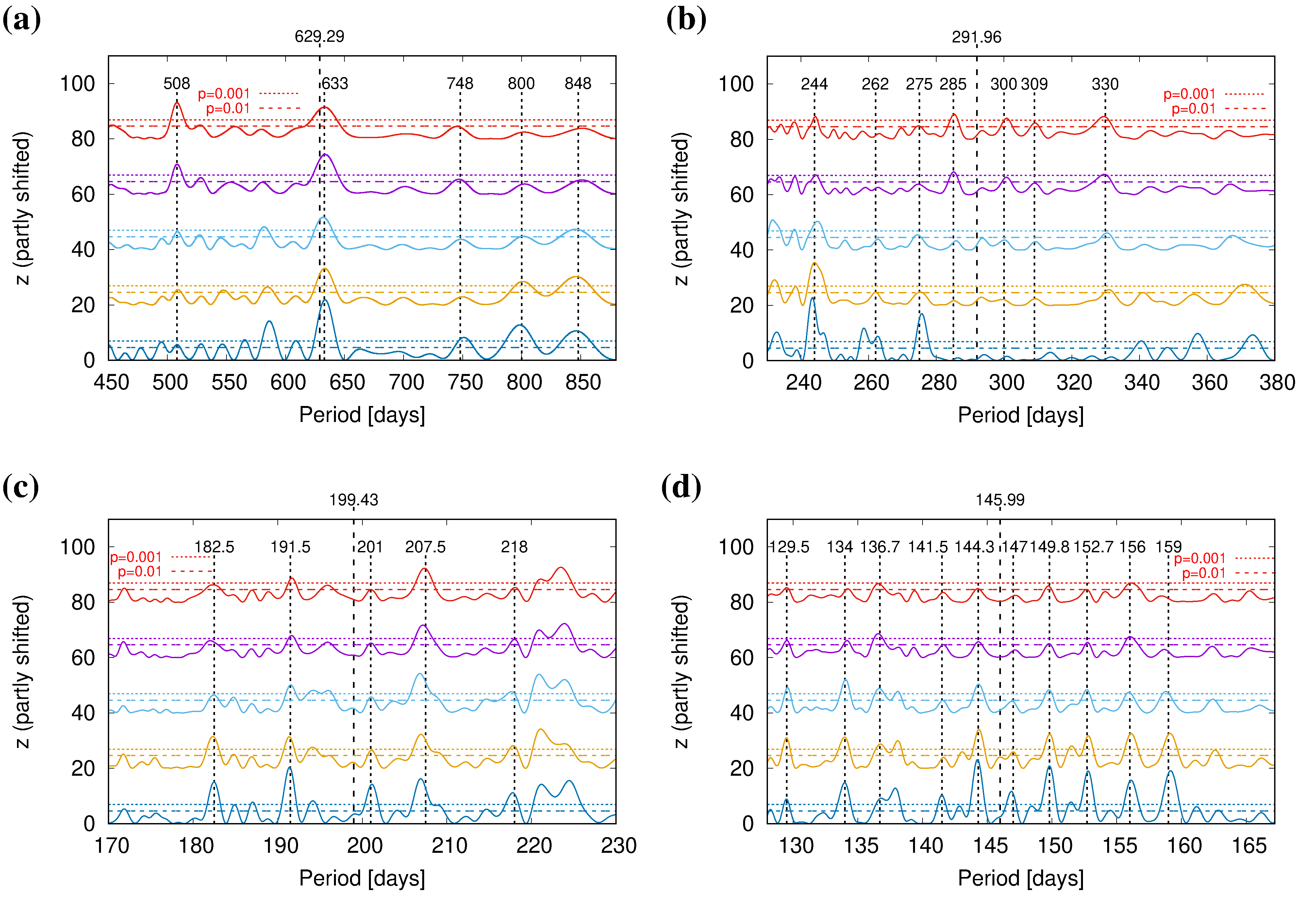}
  \caption{Curves of the $z$-values in the vicinity of $P_{629}$ (a), $P_{292}$ (b),
  $P_{199}$ (c) and $P_{146}$ (d). Again, the curves for the various
  X-flare thresholds are stacked on top of each other and separated by 20, and  the 
  dashed and dotted 
  color curves indicate $p-$values of 0.01 and 0.001, respectively. 
  The numbers above the upper abscissa 
  denote the theoretical values; the numbers within the panels denote various observed 
  peaks which may correspond to the respective central 
  peak or its upper and lower side peaks. The colors of the 
  curves are the same as in Figure 1.}
\end{figure}

Still, it is instructive to have a closer look at the upper and lower 
side bands of $P_{629}$, 
where also peaks with high $z$-values are observed
(since ``upper'' and ``lower'' refer not to the period but to 
the frequency, in our plots these side bands are situated to the left 
and right of $P_{629}$, respectively).
In doing so, we should be aware 
of the simultaneous presence of the long-term periods 
at $P_H$, $P_S$, $P^*_I$, $P^*_T$, and $P^*_Q$  as shown in Figure 1(b).
In principle, high amplitudes of these long-term signals can result 
in the side peaks of the short-term signals becoming higher than the 
very central peak.
Moreover, we should be aware that asymmetries in the amplitudes 
of the upper and lower side peaks may occur in case that the
amplitude and the phase are modulated differently. This effect, 
in the limiting case 
also known as {\it single-sideband modulation}, will be illustrated in 
Appendix A.

In any case, a word of warning is in order here:
although in all following analyses we will
try to find the most plausible attribution of the respective 
side peaks, we cannot exclude 
that some of them have a different origin.

Starting with Figure 2(a), we observe a notable
upper side peak at $P^*_{u,T} = 508$, which is particularly strong 
for the higher X-flare thresholds X10 and X6.
As can be seen in Table 1, this observed period $P^*_{u,T}$ 
corresponds quite well to the theoretically predicted value 
$P_{u,T} = (P^{-1}_{629} + P_T^{*-1})^{-1} = 505.5$ (it would correspond 
even better to 
 $P_{u,T} = (P^{*-1}_{629} + P_T^{*-1})^{-1} =507.9$, 
where $P_{629}$ is replaced by the observed period of $P^*_{629} = 633$).

Then, let us turn to the lower side band, situated to the right of $P_{629}$.
Here, we first observe one peak at $P^*_{l,S}=748$\,days, which is very close 
to the theoretical value 
$P_{l,S}=(P^{-1}_{629}-P_S^{-1})^{-1}=745.3$.
Another peak, denoted by $P^*_{l,I}=800$,
is close to $P_{l,I}=(P_{629}^*-P_I^{*-1})^{-1}=786.4$.
The fact that this peak becomes dominant only
for lower X-flare thresholds is not surprising since $P^*_{I}$ 
acquires some prominence only for the
X2 and X1 thresholds (see Figure 1b). 
The third peak in the lower side band, 
observed at $P^*_{l,T}=848$\,days, is not far 
from $P_{l,T}=(P_{629}^{-1}- P_T^{*-1})^{-1}=833.3$.

Having thus identified the pair of upper ($P^*_{u,T}$) 
and the lower ($P^*_{l,T}$) side peaks
of $P_{629}$, 
we now ask if it would have been possible to recover from them
the position of the central peak if this had not 
been available by observation. 
This exercise is carried out in the lower
part of Table 1, where we determine the {\it inferred} period 
(indicated by a double asterisk)
$P_{629}^{**}=2 (P_{u,T}^{*-1}+P_{l,T}^{*-1})^{-1}=635.4$. 
Obviously, this is indeed reasonably close to the observed 
period $P^*_{629}=633$. Likewise, we also try to recover the
third harmonics $P_T$ by computing 
$P_T^{**}=2 (P_{u,T}^{*-1}-P_{l,T}^{*-1})^{-1}=2534$, 
which is again close to the observed value $P_T^{*}=2570$
(later we will learn, however, that the inference of  
those long-term periods by this method is often
inaccurate due to the subtraction of two
nearly equal numbers).

To summarize this part on the QBO-period $P_{629}$:
this peak and various of its side peaks agree 
quite accurately with the theoretical predictions.
The amplitudes of the side peaks can become comparable to that of 
the central peak. However, their relative strengths may depend on 
the X-flare threshold used, and some of them might even be completely absent.

Let us now turn to the Venus-Earth spring tide period  
$P_{292}=291.96$\,days.
As was already seen in Figure 1(c), peaks in this region
are less pronounced than those at or around
$P_{629}$. In particular, there seems
to be no significant peak at the very central period of 292\,days, as 
shown in Figure 2(b). But what about side peaks?
In Table 2 we make an attempt to identify the upper and lower
side peaks whose values are indicated in Figure 2(b). 
We will do this by focusing on pairs of peaks. 
The first pair, with the observed periods 
$P^*_{u,T}=262$ and $P^*_{l,T}=330$, agrees
quite well with the theoretical pair
$P_{u,T}=(P_{292}^{-1}+P_T^{*-1})^{-1}=262.2$
and 
$P_{l,T}=(P_{292}^{-1}-P_T^{*-1})^{-1}=329.4$.
Similarly, the second pair, consisting of $P^*_{u,S}=275$ and 
$P^*_{l,S}=309$, may be related to the theoretical pair
$P_{u,S}=(P_{292}^{-1}+P_S^{-1})^{-1}=272.3$
and 
$P_{l,S}=(P_{292}^{-1}-P_S^{-1})^{-1}=314.9$.
Finally, the  third pair $P^*_{u,H}=285$ and $P^*_{l,H}=300$, 
may correspond to the theoretical pair
$P_{u,H}=(P_{292}^{-1}+P_H^{-1})^{-1}=281.8$
and 
$P_{l,H}=(P_{292}^{-1}-P_H^{-1})^{-1}=302.9$.
At present, we have no explanation for another peak at 244.

\begin{table}[h]
\scriptsize
\caption{Same as Table 1, but for $P_{292}=291.96$\,days.}
\begin{tabular}{rrrr}
\hline
\hline
Observed side peaks&&Inferred from theory  \\
\hline
$P^*_{u,T}=262$&&$P_{u,T}=(P_{292}^{-1}+P_T^{*-1})^{-1}=262.2$\\
$P^*_{l,T}=330$&&$P_{l,T}=(P_{292}^{-1}-P_T^{*-1})^{-1}=329.4$\\
$P^*_{u,S}=275$&&$P_{u,S}=(P_{292}^{-1}+P_S^{-1})^{-1}=272.3$\\
$P^*_{l,S}=309$&&$P_{l,S}=(P_{292}^{-1}-P_S^{-1})^{-1}=314.7$\\
$P^*_{u,H}=285$&&$P_{u,H}=(P_{292}^{-1}+P_H^{-1})^{-1}=281.8$\\
$P^*_{l,H}=300$&&$P_{l,H}=(P_{292}^{-1}-P_H^{-1})^{-1}=302.9$\\
\hline
Theory/Observed && Inferred from observed side peaks\\
\hline
$P_{292}=292.0$&&$P_{292}^{**}=2 (P_{u,T}^{*-1}+P_{l,T}^{*-1})^{-1}=292.1$\\
$P_T^*=2570$&&$P_T^{**}=2 (P_{u,T}^{*-1}-P_{l,T}^{*-1})^{-1}=2543$\\
$P_{292}=292.0$&&$P_{292}^{**}=2 (P_{u,S}^{*-1}+P_{l,S}^{*-1})^{-1}=291.1$\\
$P_S=4043$&&$P_S^{**}=2 (P_{u,S}^{*-1}-P_{l,S}^{*-1})^{-1}=4998$\\
$P_{292}=292.0$&&$P_{292}^{**}=2 (P_{u,H}^{*-1}+P_{l,H}^{*-1})^{-1}=292.3$\\
$P_H=8043$&&$P_H^{**}=2 (P_{u,H}^{*-1}-P_{l,H}^{*-1})^{-1}=11400$\\
\hline
\end{tabular}
\end{table}

As in the case of $P_{629}$, we can now again try to 
recover the position of the central peak 
from  those three pairs of observed side peaks. 
The three results for the inferred 
$P^{**}_{292}$ are, as shown on the lower part of Table 2, 
quite reasonably close to the theoretical 
value $P_{292}=291.96$. 
Not surprisingly, though, the corresponding values  inferred for the long periods 
$P^{**}_T$,
$P^{**}_S$, and $P^{**}_H$ deteriorate with increasing
period length.

To summarize this part: Compared with $P_{629}$, there is no 
significant central peak close to the Venus-Earth spring tide period $P_{292}$.
Yet, the analysis of the three pairs of observed upper and lower side peaks 
provides very good inferences of the position of the central peak.

In the next step we analyze the Earth-Jupiter spring tide at $P_{199}=199.43$\,days.
As shown in Figure 2(c), there is a nearby peak at 201\,days, whose $z$-value increases with decreasing X-flare threshold.
It is, however, not clear if this indeed 
represents the central peak.
As for the side peaks, it seems highly plausible to identify the 
observed pair $P^*_{u,T}=182.5$ and $P^*_{l,T}=218$ with the theoretical pair
$P_{u,T}=(P_{199}^{-1}+P_T^{*-1})^{-1}=185.0$
and 
$P_{l,T}=(P_{199}^{-1}-P_T^{*-1})^{-1}=216.2$, and the observed
pair $P^*_{u,S}=191.5$ and $P^*_{l,S}=207.5$ with the theoretical pair
$P_{u,S}=(P_{199}^{-1}+P_S^{-1})^{-1}=190.0$
and 
$P_{l,S}=(P_{199}^{-1}-P_S^{-1})^{-1}=209.7$.

When inferring from those two pairs the central peak we recover
$P^{**}_{199}=198.7$ and $P^{**}_{199}=199.2$ which both are very
reasonable.

\begin{table}[h]
\scriptsize
\caption{Same as Table 1, but for $P_{199}=199.43$\,days.}
\begin{tabular}{rrrr}
\hline
\hline
Observed side peaks&&Inferred from theory  \\
\hline
$P^*_{u,T}=182.5$&&$P_{u,T}=(P_{199}^{-1}+P_T^{*-1})^{-1}=185.0$\\
$P^*_{l,T}=218$&&$P_{l,T}=(P_{199}^{-1}-P_T^{*-1})^{-1}=216.2$\\
$P^*_{u,S}=191.5$&&$P_{u,S}=(P_{199}^{-1}+P_S^{-1})^{-1}=190.0$\\
$P^*_{l,S}=207.5$&&$P_{l,S}=(P_{199}^{-1}-P_S^{-1})^{-1}=209.7$\\
\hline
Theory/Observed && Inferred from observed side peaks\\
\hline
$P_{199}=199.4$&&$P_{199}^{**}=2 (P_{u,T}^{*-1}+P_{l,T}^{*-1})^{-1}=198.7$\\
$P_T^*=2570$&&$P_T^{**}=2 (P_{u,T}^{*-1}-P_{l,T}^{*-1})^{-1}=2241$\\
$P_{199}=199.4$&&$P_{199}^{**}=2 (P_{u,S}^{*-1}+P_{l,S}^{*-1})^{-1}=199.2$\\
$P_S=4043$&&$P_S^{**}=2 (P_{u,S}^{*-1}-P_{l,S}^{*-1})^{-1}=4967$\\
\hline
\end{tabular}
\end{table}

We now turn to the important period $P_{146}=145.99$\,days which is 
the closest one to the period of 154\,days as originally found by \cite{Rieger1984}.
As shown in Figure 2(d), a quite complicated picture of peaks 
emerges in this region.
Basically, we observe eight quite robust peaks between
134 and 159\,days, which become stronger when decreasing the X-flare threshold. Interestingly, when lowering this threshold, we observe
an increasing
prevalence of lower side peaks.
For the dark blue curve (X1+GLE) one could even guess
that these peaks are accumulated close to the 
154 days as observed by \cite{Rieger1984}.

\begin{table}[h]
\scriptsize
\caption{Same as Table 1, but for $P_{146}=145.99$\,days.}
\begin{tabular}{rrrr}
\hline
\hline
Observed side peaks&&Inferred from theory  \\
\hline
$P^*_{u,Q}=134$&&$P_{u,Q}=(P_{146}^{-1}+P_Q^{*-1})^{-1}=136.5$\\
$P^*_{l,Q}=159$&&$P_{l,Q}=(P_{146}^{-1}-P_Q^{*-1})^{-1}=156.8$\\
$P^*_{u,T}=136.7$&&$P_{u,T}=(P_{146}^{-1}+P_T^{*-1})^{-1}=138.2$\\
$P^*_{l,T}=156$&&$P_{l,T}=(P_{146}^{-1}-P_T^{*-1})^{-1}=154.8$\\
$P^*_{u,S}=141.5$&&$P_{u,S}=(P_{146}^{-1}+P_S^{-1})^{-1}=140.9$\\
$P^*_{l,S}=152.7$&&$P_{l,S}=(P_{146}^{-1}-P_S^{-1})^{-1}=151.5$\\
$P^*_{u,H}=144.3$&&$P_{u,H}=(P_{146}^{-1}+P_H^{-1})^{-1}=143.4$\\
$P^*_{l,H}=149.8$&&$P_{l,H}=(P_{146}^{-1}-P_H^{-1})^{-1}=148.7$\\
\hline
Theory/Observed && Inferred from observed side peaks\\
\hline
$P_{146}=146.0$&&$P_{146}^{**}=2 (P_{u,Q}^{*-1}+P_{l,Q}^{*-1})^{-1}=146.0$\\
$P_Q^*=2100$&&$P_Q^{**}=2 (P_{u,Q}^{*-1}-P_{l,Q}^{*-1})^{-1}=2108$\\
$P_{146}=146.0$&&$P_{146}^{**}=2 (P_{u,T}^{*-1}+P_{l,T}^{*-1})^{-1}=145.7$\\
$P_T^*=2570$&&$P_T^{**}=2 (P_{u,T}^{*-1}-P_{l,T}^{*-1})^{-1}=2210$\\
$P_{146}=146.0$&&$P_{146}^{**}=2 (P_{u,S}^{*-1}+P_{l,S}^{*-1})^{-1}=146.9$\\
$P_S=4043$&&$P_S^{**}=2 (P_{u,S}^{*-1}-P_{l,S}^{*-1})^{-1}=3858$\\
$P_{146}=146.0$&&$P_{146}^{**}=2 (P_{u,H}^{*-1}+P_{l,H}^{*-1})^{-1}=147.0$\\
$P_H=8086$&&$P_H^{**}=2 (P_{u,H}^{*-1}-P_{l,H}^{*-1})^{-1}=7860$\\
\hline
\end{tabular}
\end{table}

In Table 4, we try to identify the eight main upper and lower side peaks. 
While this attribution is not perfectly unique,
the pairwise ordering in upper and lower side peaks 
looks quite sensible, 
in particular since the corresponding recovery $P^{**}_{146}$ 
of the central peak 
(as shown in the lower part of Table 4) provides - for all four pairs
of upper and lower side peaks - values which are consistently 
close to the theoretical value of $P_{146}=146$.

We are thus led to the preliminary 
conclusion that the original Rieger peak at 
154\,days may represent just an average  over the lower side band peaks 
lying to the right of our theoretical central peak  $P_{146}$. The prevalence of the 
lower side band is indeed compatible with \cite{Rieger1984} who also used
a large number of events (i.e., 139 events 
within only 3.5 years). We again refer to Appendix A where, 
for the representative period $P_{146}$, this prevalence 
is shown to result from different modulations of amplitude and phase.

\subsection{Sub-Rieger periods}
After having discussed the signals from the QBO down to the Rieger period, 
in the following we will consider the five predicted
periods which are shorter than the Rieger period, namely 
the Venus-Jupiter spring tide with  $P_{118}=118.50$\,days, half of the 
Earth-Jupiter spring tide with $P_{100}=99.72$\,days, 
the two periods $P_{84}=84.29$\,days 
and $P_{74}=74.33$\,days, and half of the Venus-Jupiter spring tide 
with $P_{59}=59.25$\,days. Interestingly, peaks close to the latter three values 
showed up
already in Figure 2(a) of \cite{Rieger1984}. We will come back to 
them further below.

\begin{figure}[t]
  \centering
\includegraphics[width=0.99\textwidth]{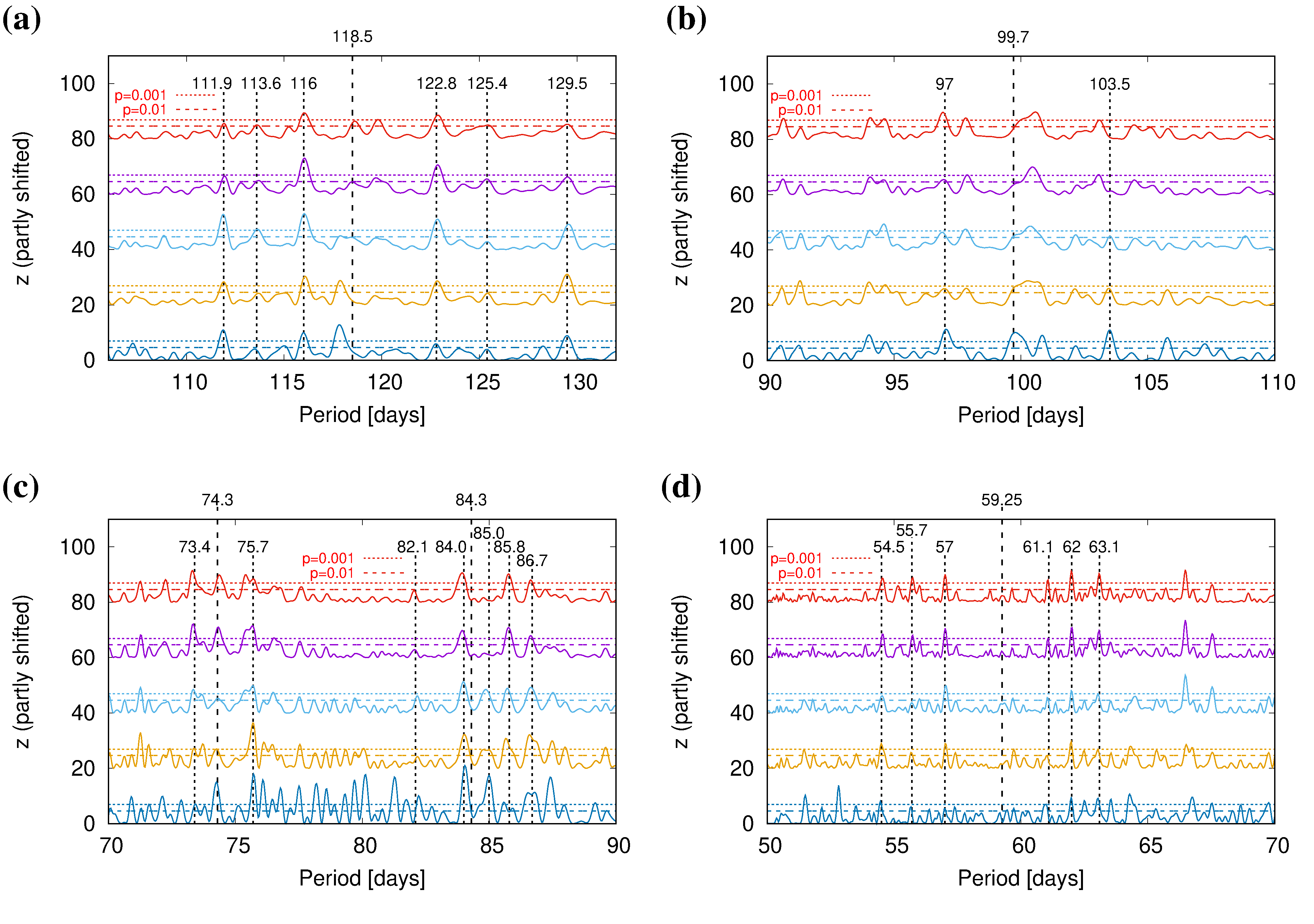}
  \caption{Same as Figure 2 but for $P_{118}$ (a), $P_{100}$ (b), $P_{84}$ and $P_{74}$ (c), and $P_{59}$ (d).}
\end{figure}

\begin{table}[h]
\scriptsize
\caption{Same as Table 1, but for $P_{118}=118.5$\,days.}
\begin{tabular}{rrrr}
\hline
\hline
Observed side peaks&&Inferred from theory \\
\hline
$P^*_{u,Q}=111.9$&&$P_{u,Q}=(P_{118}^{-1}+P_Q^{*-1})^{-1}=112.2$\\
$P^*_{u,T}=113.6$&&$P_{u,T}=(P_{118}^{-1}+P_T^{*-1})^{-1}=113.3$\\
$P^*_{l,T}=125.4$&&$P_{l,T}=(P_{118}^{-1}-P_T^{*-1})^{-1}=124.2$\\
$P^*_{u,S}=116.0$&&$P_{u,S}=(P_{118}^{-1}+P_S^{-1})^{-1}=115.1$\\
$P^*_{l,S}=122.8$&&$P_{l,S}=(P_{118}^{-1}-P_S^{-1})^{-1}=122.1$\\
$P^*_{l,2T}=129.5$&&$P_{l,2T}=(P_{118}^{-1}-2 P_T^{*-1})^{-1}=130.5$\\
\hline
Theory/Observed && Inferred from observed side peaks\\
\hline
$P_{118}=118.5$&&$P_{118}^{**}=2 (P_{u,T}^{*-1}+P_{l,T}^{*-1})^{-1}=119.2$\\
$P_T^*=2570$&&$P_T^{**}=2 (P_{u,T}^{*-1}-P_{l,T}^{*-1})^{-1}=2414$\\
$P_{118}=118.5$&&$P_{118}^{**}=2 (P_{u,S}^{*-1}+P_{l,S}^{*-1})^{-1}=119.3$\\
$P_S=4043$&&$P_S^{**}=2 (P_{u,S}^{*-1}-P_{l,S}^{*-1})^{-1}=3858$\\
\hline
\end{tabular}
\end{table}

Let us begin with $P_{118}=118.5$ which is illustrated in Figure 3(a). 
While in Figure 2(a) of \cite{Rieger1984} there were no peaks 
around this value,
quite interesting features close to it were indeed shown 
in Figures 2 and 3 of 
\cite{Gurgenashvili2021}.

Obviously, in Figure 3(a)  peaks close to 118.5 are only 
weakly expressed for higher X-flare thresholds, and for lower thresholds 
they shift slightly to the left.
Yet, there are two stable pairs of upper/lower side band peaks 
at 113.6/125.4  and 116.0/122.8 which fit reasonably well
to the corresponding theoretical values. Another upper side 
peak at 111.9 might correspond to
$P_{u,Q}=(P_{118}^{-1}+P_Q^{*-1})^{-1}=112.2$.
The lower side peak at 129.5, however, does not fit easily into this scheme.
In Table 5 we indicate the possibility that this might be a {\it secondary}
side peak, denoted by $P^*_{l,2T}$, although this interpretation 
is a bit speculative.
At any rate, the lower part of Table 5 shows that the 
central peak that would be inferred from the observed side peaks 
is indeed not too far from $P_{118}=118.5$.

The next period, $P_{100}=99.72$, is only well expressed for 
the X1 threshold, while it 
sits at slightly higher values for higher X-flare thresholds.
The first upper side peak to the left, at 97, fits well to
the theoretical prediction 
$P_{u,S}=(P_{100}^{-1}+P_S^{-1})^{-1}=97.3$, while
the first lower side peak, at 103.5, fits  to
$P_{l,T}=(P_{100}^{-1}-P_T^{*-1})^{-1}=103.7$.
Since, unfortunately, these two values do not represent an upper/lower side 
peak pair, we refrain
here from trying to recover the central peak.

\begin{table}[h]
\scriptsize
\caption{Same as Table 1, but for $P_{100}=99.72$\,days.}
\begin{tabular}{rrrr}
\hline
\hline
Observed side peaks&&Inferred from theory  \\
\hline
$P^*_{u,S}=97$&&$P_{u,S}=(P_{100}^{-1}+P_S^{-1})^{-1}=97.3$\\
$P^*_{l,T}=103.5$&&$P_{l,T}=(P_{100}^{-1}-P_T^*)^{-1}=103.7$\\
\hline
\end{tabular}
\end{table}

We now turn to the two periods $P_{84}=84.29$ 
and $P_{74}=74.33$ which are jointly illustrated in Figure 3(c). 
Remarkably, two peaks very close to these values already showed 
up in Figure 2 of \cite{Rieger1984}.

For $P_{84}$ the situation around the central peak is again not 
perfectly unique. The observed value might be either the peak at $P_{84}^*=84.0$
or at $P_{84}^*=85.0$. Still, the central peak inferred from the observed pair
$P^*_{u,S}$ and $P^*_{l,S}$ appears quite reasonable.
 
\begin{table}[h]
\scriptsize
\caption{Same as Table 1, but for $P_{84}=84.29$\,days.}
\begin{tabular}{rrrr}
\hline
\hline
Observed side peaks&&Inferred from theory   \\
\hline
$P^*_{u,S}=82.1$&&$P_{u,S}=(P_{84}^{-1}+P_S^{-1})^{-1}=82.6$\\
$P^*_{l,S}=85.8$&&$P_{l,S}=(P_{84}^{-1}-P_S^{-1})^{-1}=86.1$\\
$P^*_{l,H}=85.0$&&$P_{l,H}=(P_{84}^{-1}-P_H^{-1})^{-1}=85.2$\\
$P^*_{l,T}=86.7$&&$P_{l,T}=(P_{84}^{-1}-P_T^{*-1})^{-1}=87.2$\\
\hline
Theory/Observed && Inferred from observed side peaks\\
\hline
$P_{84}=84.3$&&$P_{84}^{**}=2 (P_{u,S}^{*-1}+P_{l,S}^{*-1})^{-1}=83.9$\\
$P_S=4043$&&$P_S^{**}=2 (P_{u,S}^{*-1}-P_{l,S}^{*-1})^{-1}=3807$\\
\hline
\end{tabular}
\end{table}

Much clearer seems to be the situation for $P_{74}=74.33$. Here 
the observed position of the central peak is exactly the same, while the 
two side peaks, $P^*_{u,S}=73.4$ and $P^*_{l,S}=75.7$, 
seem to correspond to the theoretical
values
$P_{u,S}=(P_{74}^{-1}+P_S^{-1})^{-1}=73.0$ and
$P_{l,S}=(P_{74}^{-1}-P_S^{-1})^{-1}=75.7$.
The inferred values are then 
$P_{74}^{**}=2 (P_{u,S}^{*-1}+P_{l,S}^{*-1})^{-1}=74.5$ 
for the central peak (which is quite accurate), and
$P_S^{**}=2 (P_{u,S}^{*-1}-P_{l,S}^{*-1})^{-1}=4831$
for the Schwabe period which is also acceptable.

\begin{table}[h]
\scriptsize
\caption{Same as Table 1, but for $P_{74}=74.33$\,days.}
\begin{tabular}{rrrr}
\hline
\hline
Observed side peaks&&Inferred from theory \\
\hline
$P^*_{u,S}=73.4$&&$P_{u,S}=(P_{74}^{-1}+P_S^{-1})^{-1}=73.0$\\
$P^*_{l,S}=75.7$&&$P_{l,S}=(P_{74}^{-1}-P_S^{-1})^{-1}=75.7$\\
\hline
Theory/Observed && Inferred from observed side peaks\\
\hline
$P_{74}=74.3$&&$P_{74}^{**}=2 (P_{u,S}^{*-1}+P_{l,S}^{*-1})^{-1}=74.5$\\
$P_S^=4043$&&$P_S^{**}=2 (P_{u,S}^{*-1}-P_{l,S}^{*-1})^{-1}=4831$\\
\hline
\end{tabular}
\end{table}

Finally, in Figure 3(d) we analyze the peaks around $P_{59}=59.25$.
Obviously, there are no significant features close to this value.
However, we can identify three pairs of upper/lower side peaks which are 
analyzed in Table 9.
The observed pair closest to the central peak, i.e. 57.0/61.1, may 
correspond to $P_{u,T}=57.9$ and $P_{l,T}=60.6$.
But what about the other two pairs, i.e. 55.7/62.0 and 54.5/63.1?
One possibility, illustrated in Table 9, is to interpret them as
pairs of {\it secondary} and {\it tertiary} side peaks, respectively.
While this interpretation is, again, a bit speculative, the central
peaks and the $P^*_T$ as inferred in the lower part of Table 9 
appear reasonable.

\begin{table}[h]
\scriptsize
\caption{Same as Table 1, but for $P_{59}=59.25$\,days.}
\begin{tabular}{rrrr}
\hline
\hline
Observed side peaks&&Inferred from theory  \\
\hline
$P^*_{u,T}=57.0$&&$P_{u,T}=(P_{59}^{-1}+P_T^{*-1})^{-1}=57.9$\\
$P^*_{l,T}=61.1$&&$P_{l,T}=(P_{59}^{-1}-P_T^{*-1})^{-1}=60.6$\\
$P^*_{u,2T}=55.7$&&$P_{u,2T}=(P_{59}^{-1}+2 P_T^{*-1})^{-1}=56.6$\\
$P^*_{l,2T}=62.0$&&$P_{l,2T}=(P_{59}^{-1}-2 P_T^{*-1})^{-1}=62.1$\\
$P^*_{u,3T}=54.5$&&$P_{u,3T}=(P_{59}^{-1}+3 P_T^{*-1})^{-1}=55.4$\\
$P^*_{l,3T}=63.1$&&$P_{l,3T}=(P_{59}^{-1}-3 P_T^{*-1})^{-1}=63.6$\\
\hline
Theory/Observed && Inferred from observed side peaks\\
\hline
$P_{59}=59.2$&&$P_{59}^{**}=2 (P_{u,T}^{*-1}+P_{l,T}^{*-1})^{-1}=59.0$\\
$P_T^*=2570$&&$P_T^{**}=2 (P_{u,T}^{*-1}-P_{l,T}^{*-1})^{-1}=1699$\\
$P_{59}=59.2$&&$P_{59}^{**}=2 (P_{u,2T}^{*-1}+P_{l,2T}^{*-1})^{-1}=58.7$\\
$P_T^*=2570$&&$P_T^{**}=4 (P_{u,2T}^{*-1}-P_{l,2T}^{*-1})^{-1}=2192$\\
$P_{59}=59.2$&&$P_{59}^{**}=2 (P_{u,3T}^{*-1}+P_{l,3T}^{*-1})^{-1}=58.5$\\
$P_T^*=2570$&&$P_T^{**}=6 (P_{u,3T}^{*-1}-P_{l,3T}^{*-1})^{-1}=2399$\\
\hline
\end{tabular}
\end{table}

\section{Conclusion}
Motivated by the remarkable agreement between the QBO period inferred from observations and that predicted by our model of the tidally synchronized solar dynamo, this paper addresses the central question of whether the remaining eight periods resulting from magneto-Rossby wave theory can also be found in 
solar activity data.
In this endeavor, we were further encouraged by the fact that 
the original spectra of \cite{Rieger1984} not only contained
the famous Rieger period of 154 days (which is close to our $P_{146}$), but 
also clear peaks very close to our values of $P_{84}$ and $P_{74}$.

To validate phase stability over several solar cycles, we 
applied the Rayleigh test to long-term data sets of strong solar activity, 
including the 72 GLE events since 1956, as well as X-class flare data since 1976
with varying thresholds.
As expected, the spectra of these data contain strong peaks at the Schwabe 
cycle and the third harmonic of the Hale cycle. However, they partly also 
contain the Hale cycle itself and two other periods.
In the presence of these long-term periodicities, the nine 
short-term peaks predicted by theory between 59 and 629 days 
were always accompanied, and often dominated, by upper and 
lower side bands.

For each of the nine periods, we tried to determine the central peak 
(if identifiable) as well as the observed upper and lower side peaks. 
Utilizing the latter, we also cross-checked if the central peak that 
would result from them
is close to the theoretical one. 

In general, we found plausible interpretations of the observed side peaks for all nine theoretical periods, although with some differences between very accurate matchings, particularly for $P_{74}, P_{84}$ and $P_{629}$, and some less convincing cases, such as $P_{292}$ and $P_{100}$.
For the Rieger-type period $P_{146}$ we found a significant number of side peaks with partly very high z-values. The growing prevalence of lower side peaks when lowering the X-flare threshold may explain the value of 154 days, as originally found by 
\cite{Rieger1984}, while we still consider the central peak 
at 146\,days to be quite plausible.

We reiterate the caveat that not all attributions 
of the individual side peaks may be completely correct. 
Generally, however, the significant amplitudes of the $z-$values around 
the predicted nine periods (see Figures 1(c,d)) 
lend further weight to the hypothesis that solar activity is 
largely determined by the action of tidally triggered 
magneto-Rossby waves.

\section*{Appendix A}

In this Appendix, we demonstrate how an asymmetry between the upper and lower side peaks can arise in case that the amplitude and the phase are modulated differently. Inspired by observations related to the Rieger period (see Figure 2(d)), we consider a simplified synthetic model comprising one carrier period $T_{146}=146$\,days and a modulation period of 
$T_S=4043$\,days. 

Specifically,  we study the two signals
\begin{eqnarray}
s_{in}(t)&=&\frac{1}{4}\left[1+ \cos(2 \pi t/T_S) \right]
\left[1+ \cos(2 \pi (t/T_{146}+c \cos(2 \pi t/T_S))) \right]
\end{eqnarray}
and 
\begin{eqnarray}
s_{out}(t)&=&\frac{1}{4}\left[1+ \cos(2 \pi t/T_S) \right]
\left[1+ \cos(2 \pi (t/T_{146}+c \sin(2 \pi t/T_S))) \right] \; .
\end{eqnarray}

Both of them contain an amplitude modulation with $T_S$ as the first term. However, their difference lies in the phase modulation in the second term, which is {\it in-phase} 
for $s_{in}(t)$ (using the cosine, as in the amplitude modulation term) and {\it out-of-phase} 
for $s_{out}(t)$ (using the sine). The constant $c$ is a measure of how much the maxima of the carrier signal vary between earlier and later times. 

Figure 4(a) shows $s_{in}(t)$ for three choices of the constant $c$ in the restricted interval between 2000 and 6100 days, which covers  one Schwabe cycle. In the central, high-amplitude segment around the day 4043, a negative value of $c$ leads to a shift of the curve towards later times everywhere, while a positive value shifts the curve to earlier times.

We now generate synthetic data sets of ``extreme events'' for three choices of $c$.  To do this, we first demand the instantaneous value of $s_{in}(t)$ to be greater than 0.5. Secondly, to introduce some stochasticity, we randomly select only every tenth day on average. Using a total duration of 25,000 days, the number of ``extreme events'' thus obtained amounts to around 500, comparable to the X1+GLE dataset shown in Figure 1. The colored open circles in Figure 4(a) show the results for the corresponding values of $c$. Clearly,  ``extreme events'' are only observed when $s_{in}>0.5$, but they also contain a stochastic element. A systematic motion to the left and right is visible for $c = 0.25$ and $c = -0.25$, respectively.

Figure 4(b) now shows the $z$-value resulting from the Rayleigh test when using these ``extreme events'' as input. As expected, for $c = 0$, a dominant central peak appears at the carrier period $T_{146}$ , accompanied by smaller peaks in the lower and upper side bands. Interestingly, this pattern barely changes when $c$ is set to 0.25 or -0.25. Apart from some slight changes due to the stochastic choices, the amplitude ratios of the peaks remain approximately the same.  We therefore conclude that in-phase modulation of 
amplitude and phase  (both using the cosine function) leads to symmetric sidebands.

This behavior changes when switching to out-of-phase modulation of amplitude and phase (one with the cosine and the other with the sine). Figure 4(c) shows the curves $s_{out}(t)$, again for three values of $c$. They now shift symmetrically to the left and right, when seen from the maximum at day 4043. Overall, for $c = 0.25$ (red curve), this leads to {\it compression} of the positions of the local maxima, while {\it rarefaction} is observed for $c = -0.25$ (green curve). This effect is also visible in the resulting ``extreme events''.

The consequences of this compression or rarefaction for the z-values of the Rayleigh test are illustrated in Figure 4(d). Clearly, compression (for $c=0.25$) shifts the peaks 
to the left (i.e. into the upper sideband), whereas rarefaction (with $c=-0.25$) shifts them into the lower sideband. This is indeed analogous 
to the behavior observed in Figure 2(d).

\begin{figure}[h]
  \centering
\includegraphics[width=0.95\textwidth]{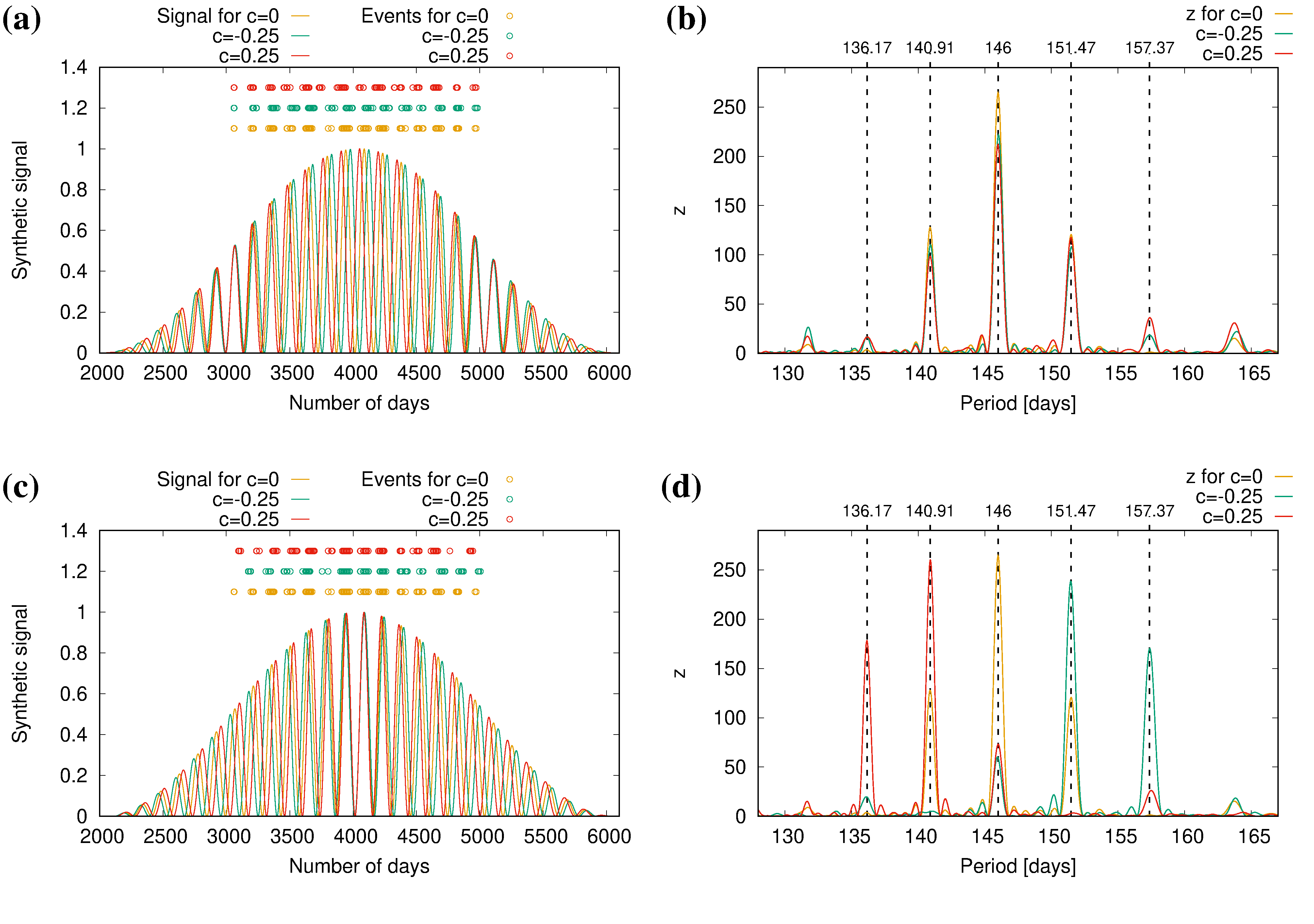}
  \caption{  Synthetic curves $s(t)$ generated with carrier and modulation periods of $P_{146}$ and $P_S$, respectively, for both in-phase (a) and out-of-phase (c) modulation of amplitude and phase. The modulus of $c$ measures the swing of the phase modulation. For each value of $c$, the corresponding ``extreme events'' (open circles) occur when $s(t)>0.5$ is demanded and then, on average, only every tenth day is selected at random. The $z$-values of the Rayleigh test resulting from these events are shown in (b) for in-phase modulation and in (d) for out-of-phase modulation.  Only (d) shows a clear symmetry breaking between the upper and lower side peaks.
  }
\end{figure}

While all this sounds quite plausible from a {\it mathematical} point of view, the question remains: what  {\it physical} effect could be responsible for the out-of-phase modulation? One possibility is that, while the rise speed of the flux tube is a monotonically increasing function of the field amplitude, the response of the magneto-Rossby waves has a maximum at a certain field strength and then decreases. Yet, a thorough investigation of this effect is beyond the scope of this paper.

\section*{Acknowledgments} 
We thank Eka Gurgenashvili for her help in providing the X-flare data.
F. S. would like to express his gratitude to the late Uwe Harlander for 
many inspiring discussions on the role of harmonic forcings and 
Rossby waves in geo- and astrophysics.

\bibliography{glesflare}{}
\bibliographystyle{aasjournal}



\end{document}